# Single Photon Emission from Deep Level Defects in Monolayer $WSe_2$

Yanxia Ye, Xiuming Dou*, Kun Ding, Yu Chen, Desheng Jiang, Fuhua Yang, and Baoquan Sun*

*State Key Laboratory of Superlattices and Microstructures, Institute of Semiconductors, Chinese Academy of Sciences, Beijing 100083, China*

We report an efficient method to observe single photon emissions in monolayer $WSe_2$ by applying hydrostatic pressure. The photoluminescence peaks of typical two-dimensional (2D) excitons show a nearly identical pressure-induced blue-shift, whereas the energy of pressure-induced discrete emission lines (quantum emitters) demonstrates a pressure insensitive behavior. The decay time of these discrete line emissions is approximately 10 ns, which is at least one order longer than the lifetime of the broad localized (L) excitons. These characteristics lead to a conclusion that the excitons bound to deep level defects can be responsible for the observed single photon emissions.

*Electronic address: xmdou04@semi.ac.cn；bqsun@semi.ac.cn

Unlike luminescence centers in traditional three-dimensional semiconductors which are located away from surface, quantum emitters in atomic thin transition-metal dichalcogenide (TMDC) semiconductors, such as $WSe_2$ and hexagonal boron nitride, have been observed at the edges of the flakes.[1-9] Local strain gradients which occur at the edges are considered to modulate the electronic states of the localized excitons,[2, 9-11] resulting in spatially and spectrally isolated single photon emission. This means that strain engineering is an effective approach to obtain spatially and spectrally isolated quantum emitters in two-dimensional (2D) semiconductors. However, the uncontrollability of strains as reported restricts to better understand and explore the electronic and optical properties of discrete emitters. Therefore, *in-situ* strain tuning becomes essential for understanding the optical properties of discrete emitters.

Here, we present an effective method to give rise to single photon emission lines in monolayer $WSe_2$ and provide experimental evidence to show that the deep level defects should be responsible for the emission of single photon lines. The results are based on the application of hydrostatic pressure technique in the low temperature photoluminescence (PL) measurement of monolayer $WSe_2$. The 2D-excitons and discrete emission lines which arise during pressure application are studied in detail by the pressure-dependent PL and time-resolved PL (TR-PL). The pressure is introduced by the piezoelectric actuator-driven diamond anvil cell (DAC) device for *in-situ* pressure tuning. In addition to the pressure coefficients (PCs) of the 2D-excitons, the abnormal pressure responses of the discrete lines are observed. The mechanism of the discrete emission lines is carefully checked-by measurements before and after the pressure engineering cycles.

In our experiments, monolayer $WSe_2$ flakes were prepared by micromechanical exfoliation from a bulk $WSe_2$ (*2D Semiconductors* supplied) on a thinned $SiO_2$/Si substrate. A high pressure can be applied to the measured samples by using diamond anvil cell (DAC) device driven by a piezoelectric actuator.[12] Condensed argon was used as the pressure-transmitting medium, and the ruby $R_1$ fluorescence line shift was used to determine pressure. The calibrated temperature of the cryogenic DAC sample chamber is 20 K. The PL was collected by a 50x objective (NA: 0.35) and spectrally analyzed using a 0.5 monochromator equipped with a silicon charge-coupled device (CCD) at an excitation of tens of μW by a CW or pulse adjustable 640 nm semiconductor laser. Silicon single-photon counting modules with a time resolution of 380 ps were used for time-resolved PL and the second-order correlation function $g^{(2)}(\tau)$ measurements.

As it has been reported that the discrete PL lines in monolayer cannot been observed when the laser spot is focused on the central region of the monolayer $WSe_2$ flakes.[13-15] Instead, typical PL peaks of the 2D neutral exciton (2D-$X^0$), 2D charge exciton (2D-$X^-$) and defect-related L exciton band are observed at zero pressure, as shown in Fig. 1 (sample 1). When pressure is applied and increased, $X^0$ emission energy shows a blue-shift, (as indicated in Fig. 1 by arrows where $X^0$ peak shifts at a rate of 15 meV/GPa), the corresponding peak intensity decreases rapidly and turns to becomes too weak to detect. Moreover, accompanying with the evolution of a pressure-induced blue-shift and intensity decrease of 2D exciton emission, it is noticed that a lot of discrete emission lines emerge within the energy region of the broad L band (1.6-1.7 eV) and/or at the lower energy side ($h\nu$ below 1.6 eV). Figure 1 show the PL spectral curves measured at the pressures of 0.83, 1.51, 1.98, 2.32 and 4.4 GPa. Here, the laser spot is focused on the central region of the flake at each PL measurement, and the discrete lines emerge under high pressures. Actually, these lines are very weak and their intensity is incomparable with the L band emission when the pressure is below approximately 0.7 GPa. For $P > 0.7$ GPa, the discrete lines emerge clearly, which is accompanied with a simultaneous rapid decrease of $L_1$ and $L_2$ emission intensities. The diminished L emission implies that the pressure-induced defect states tend to trap more carriers and give rise to the discrete emission lines or an ensemble of discrete emission lines, as shown in Fig. 1 and also in Fig. 2(a). It is noticeable that as shown in Fig. 1, the broad envelope band composed of discrete lines shows a slightly red shift with increasing pressure, which will be discussed in detail latter.

According to the peak energy position relative to the L excitons, these discrete emission lines can be divided into two groups, i.e. peak energies below and above 1.6 eV. One group of peaks is located below 1.6 eV which is lower than the L broadband; another group of discrete line peaks are above 1.6 eV, i.e. their energies are nearly overlapped with the L broadband seen at low pressure, as represented by DL1 and DL2 marked in Fig. 2(a), respectively. The single photon emission characteristic of both lines are confirmed by using second-order correlation function $g^{(2)}(\tau)$, as depicted in the insets of Fig. 2(a). In addition, the time-resolved PL measurement is applied to characterize the dynamic behavior of individual spectral lines. Figs. 2(b) and (c) show the exciton decay curves of the DL1 and DL2, respectively. It clearly exhibits that DL1 has a single exponential decay with a lifetime of 10.3±0.5 ns. However, there are two decay times for the DL2, corresponding to the lifetimes of $\tau_1$=0.62±0.03 and $\tau_2$=7.2±0.3 ns, respectively. Here, an exciton decay time of a few ns of a discrete line is in agreement with the typical value for the quantum emitters in 2D materials.[3, 10] Instead, the reported decay time of the broad L excitons is shorter, from ten picoseconds[16] to hundreds picoseconds.[17] This means that the lifetime of discrete lines is about one or two order longer than the lifetime of the broad L band emission. In fact, as shown in Fig. 2(d) the decay curve of $L_2$ emission is measured at zero pressure. A lifetime of 0.57±0.03 ns of the decay curve is obtained by deconvolution of the PL curve from the instrument response function (IRF). Therefore, the time decay of DL2 may be related to two different kinds of exciton decay processes even their emission energy is nearly overlapped at low pressure. It is noted that the lifetime $\tau_1$ of DL2 has the same order as the $L_2$ exciton. Thus, the detected signal of DL2 may be partly mixed with some contribution of "residual" L band emission; the fast decay process of DL2 can be attributed to the radiative recombination of L excitons instead of quantum emitter.

To inspect the pressure-induced shifts of emission peaks 2D-$X^0$, 2D-$X^-$, and L excitons, as well as of the discrete lines, an *in-situ* tunable pressure on the 2D flake has been exerted by using a piezoelectric actuator-driven DAC device. The measured PL peak energies under high pressure show a blue-shift for the $X^0$, $X^-$, and L excitons, and the corresponding PL intensity decreases rapidly with increasing pressure, as presented in Fig. 3(a) (sample 3). The detailed experimental data of the PL peak energy as a function of pressure are plotted in Fig. 3(b), together with the data at zero pressure. All of these exciton peaks show blue-shift with increasing pressure. The linear functions are employed to fit the data and the obtained pressure coefficients (PCs) are 14.7±0.2, 13.0±0.6, 14.4±0.8 and 13.3±0.5 meV/GPa for $X^0$, $X^-$, $L_1$ and $L_2$ excitons, respectively. The PC of $X^0$ peak reflects a blue-shift rate of band edge of the direct K-K interband transition.[18-21] Here, the PC of $X^0$ at low temperature is nearly a half of the value for 2D-$X^0$ at room temperature reported elsewhere.[22] The discrepancy in the PC values obtained at different temperatures is at moment not understood. By comparing the PC of $L_1$ and $L_2$ excitons with that of $X^0$, it is found that defect-level-bound $L_1$ and $L_2$ excitons have nearly the same blue-shift rate as $X^0$, suggesting that the related defect levels should correspond to the hydrogen-like shallow impurities. Figure 3(c) shows the color-coded PL intensity and peak energy shift as a function of pressure (sample 4). It is noticeable that in Fig. 3(c), at zero pressure there is a typical PL emission of the 2D-$X^0$, 2D-$X^-$, and L excitons. When the applied pressure is increased to be higher than approximately 1.2 GPa, discrete emission lines gradually emerge up in the spectral region of L exciton broadband or below it. The spectral range of the discrete lines has a broader energy distribution. Especially, it extends to the lower energy side at higher pressure. This kind of pressure-induced behavior is contrary to the way of response for direct band excitons ($X^0$ and $X^-$) and shallow-donor-bound L excitons under pressure (see the white dashed lines for eye guideline). It is known that the shift of shallow levels follows the conduction band edge under pressure, while the deep levels have a much less pressure-induced shift and normally show a strongly sublinear change with pressure.[23, 24] Thus, based on the observed weak pressure response or red-shift of the discrete lines, together with a longer decay time, showing a characteristic of stronger bounding from the defect state, the discrete line emissions are attributed to the exciton emissions bound to the deep level defects which have a larger degree of atomic-like character and a stronger localization effect on excitons.

Accompanied with the appearance of discrete emitters, it is noted that the 2D-excitons are simultaneously weakened or even disappear, as shown in Figs. 2(a), 3(c) and 4(b). It is very similar to what was reported for the quantum emitters observed at the edge of etched holes, wrinkle on the flakes, or at the edge of flakes.[1, 10] It can be assumed to be a result of the competition between different kinds of radiative recombination. The 2D excitons and weakly bounded excitons are more easily to be trapped by strong localization centers newly created by hydrostatic pressure. Spatially, it was known that the discrete lines occur often at the edges of flakes and the regions of high strain gradient.[2, 9-11] We find that for most samples the studied 2D exciton peaks disappear under an applied pressure of ~ 0.6–1.5 GPa at the first cycle of *in situ* pressure-changing process. After releasing the pressure to zero, a remarkable change of PL lineshape is observed, as typically shown in Fig. 3(d). The change of zero pressure PL spectra measured at the same location implies an existence of the residual strains or cracks on the sample.

Here, the cracks appearing on monolayer $WSe_2$ may result in dangling bonds or the possible reconstruction of the flake edges, which can be considered to form deep levels within the energy bandgap. A similar effect was reported to occur after a high temperature annealing treatment of samples.[25] Furthermore, after several cycles of pressure-changing experiments, the cracks tend to be enlarged and the monolayer sample could be broken into fragments, as shown in Fig. 4(a) by optical microscope. To inspect the change in 2D and L exciton emissions, as well as the emergence of discrete emitters under pressure, several runs of pressure-tuning cycles are applied successively. The PL measurements after the 2$^{nd}$, 4$^{th}$ and 5$^{th}$ period of pressure-tuning cycles have been completed on the same location of the sample with *in situ* pressure tuning method, as shown in Fig. 4(b). It is demonstrated that with increasing pressure-tuning cycles the 2D-$X^0$ and 2D-$X^-$ exciton emissions become weaker and then even disappear. L broadband emission becomes also weak enough or diminished, then discrete emission lines appear and can be well recognized. We find that it is difficult to follow the particular track of single emission lines even though by *in situ* tuning pressure and the laser spot has been tried to fix-on the same location of the flake. With increasing pressure, old emission lines die off quickly and some new ones emerge. This is assumed to directly result in an obvious difference between comb-like emission spectra at different pressures, as shown in the 5 spectra taken on the 5$^{th}$ applied pressure cycle in Fig. 4(b). Comparing the amplified microimage of residual strains and cracks, it is confirmed that it is the deep level defects that give rise to the discrete emission lines.

In summary, we report an effective method to give rise to single photon emission lines in 2D $WSe_2$ material by applying hydrostatic pressure. We find that the 2D-$X^0$ and 2D-$X^-$ exciton emissions are quenched at high pressure. The pressure-induced residual strains or cracks on the sample can be responsible for the emergence of single photon emissions and the quenching of 2D exciton emissions. Based on nearly the same pressure-induced blue-shift rate for both 2D excitons and L excitons, the defect states which bound L excitons are attributed to shallow impurity levels. Whereas, the emergent discrete line emissions are demonstrated to be pressure less-sensitive. In addition, the exciton decay times of these lines are longer, to be approximately 10 ns. These characteristics reveal that excitons of discrete line emissions should be bound ~~on~~ to the deep level defects. The exceptionally large exciton g-factor of discrete lines reported in monolayer $WSe_2$ may be also related to the atomic-like character of the deep level defects.[1-3]

This work was supported by the National Key Basic Research Program of China grant 2012(b)B922304, the National Key Research and Development Program of China grant 2016YFA0301202, and the National Natural Science Foundation of China (Grant Nos. 11474275 and 11574301).

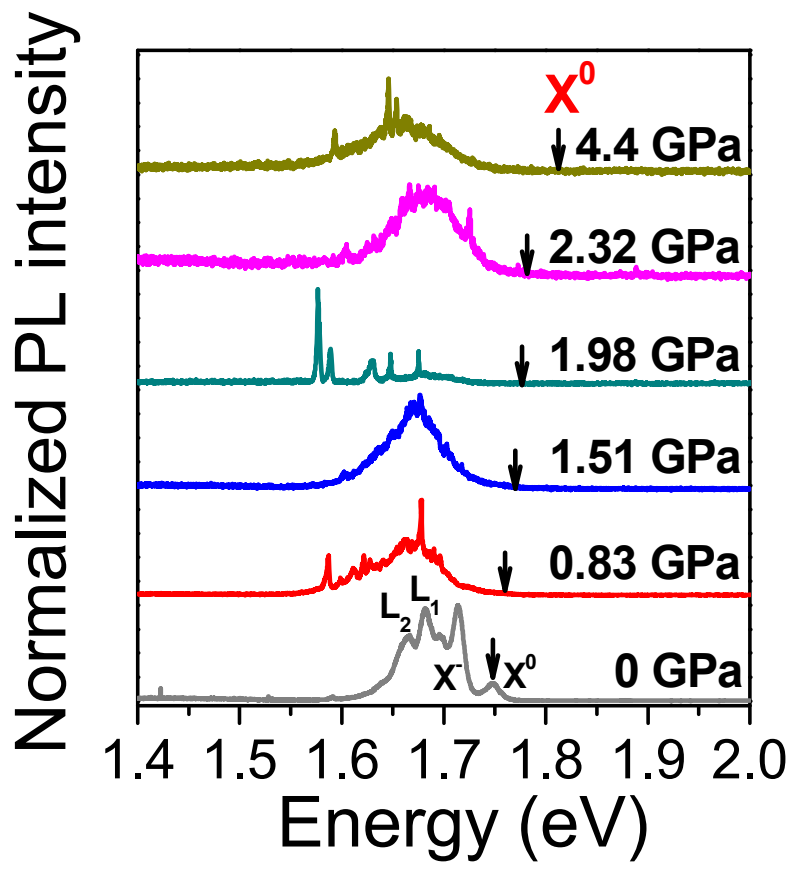

FIG. 1. Normalized PL spectra measured at the pressures of 0, 0.83, 1.51, 1.98, 2.32 and 4.4 GPa, respectively, with an excitation power of 20 μW. A typical PL spectrum at zero pressure is 2D exciton $X^0$, 2D charged exciton $X^-$ emissions and localized excitonic $L_1$ and $L_2$ emissions. The pressure response of the $X^0$ peaks is marked by black arrows according to a pressure coefficient of 14.7 meV/GPa (see Fig. 3(b)).

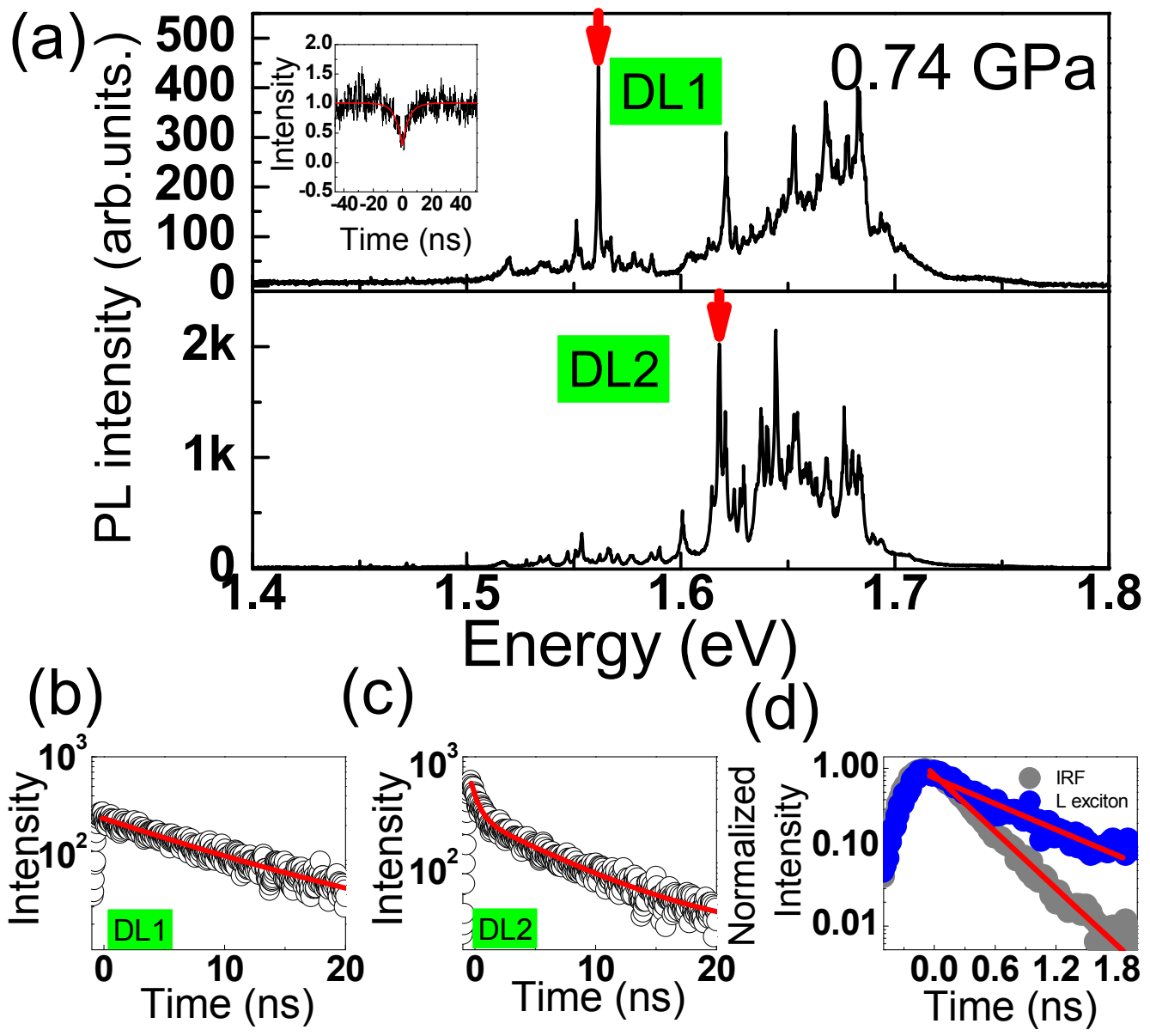

FIG. 2. (a) PL spectra of discrete line emissions at a pressure of 0.74 GPa, corresponding to two different laser excitation spots on the flake. DL1 and DL2 were used for time-resolved PL measurements. Inset: The measured $g^2(\tau)$ of DL1 corresponding $g^2(0)$ of 0.29±0.02. (b, c) Time-resolved PL spectra expressed in logarithmic coordinate, were measured by time-correlated single-photon counting (TCSPC) for DL1 and DL2, respectively. The fitting decay time is (c) 10.3±0.5 ns for DL1, (d) $\tau_1$=0.62±0.03 and $\tau_2$=7.2±0.3 ns for DL2, respectively. (d) Time-resolved PL spectra at zero pressure for $L_2$ exciton emission (blue circles, fitted as $\tau_L$=0.57±0.03ns). The gray circles correspond to the instrument response function (IRF), the fitted value is $\tau_{IRF}$=0.38±0.01ns.

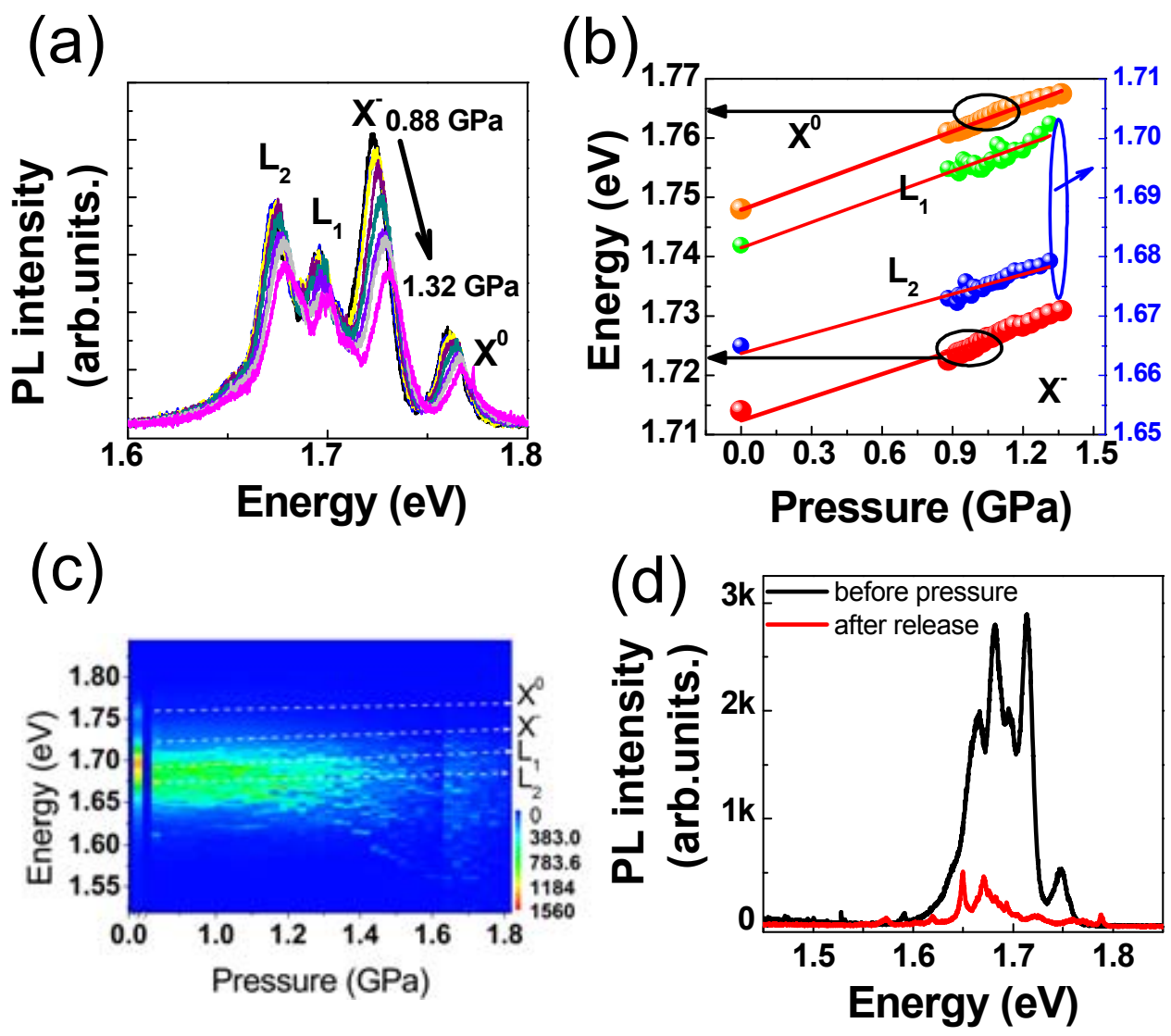

FIG. 3. (a) PL spectra response to the pressure ranging from 0.88 to 1.32 GPa for $X^0$, $X^-$, $L_1$ and $L_2$ exciton emission energies. (b) PL peak energies of $X^0$, $X^-$, $L_1$ and $L_2$ excitons as a function of pressure. The obtained pressure coefficients are 14.7±0.2, 13.0±0.6, 14.4±0.8 and 13.3±0.5 meV/GPa for $X^0$, $X^-$, $L_1$ and $L_2$ excitons, respectively. (c) Color-coded PL intensity and peak energy shift as a function of pressure. The white dashed lines are eye guidelines for the pressure response of $X^0$, $X^-$, $L_1$ and $L_2$ exciton peak energies, respectively. (d) PL spectra measured before applied pressure and after releasing the pressure.

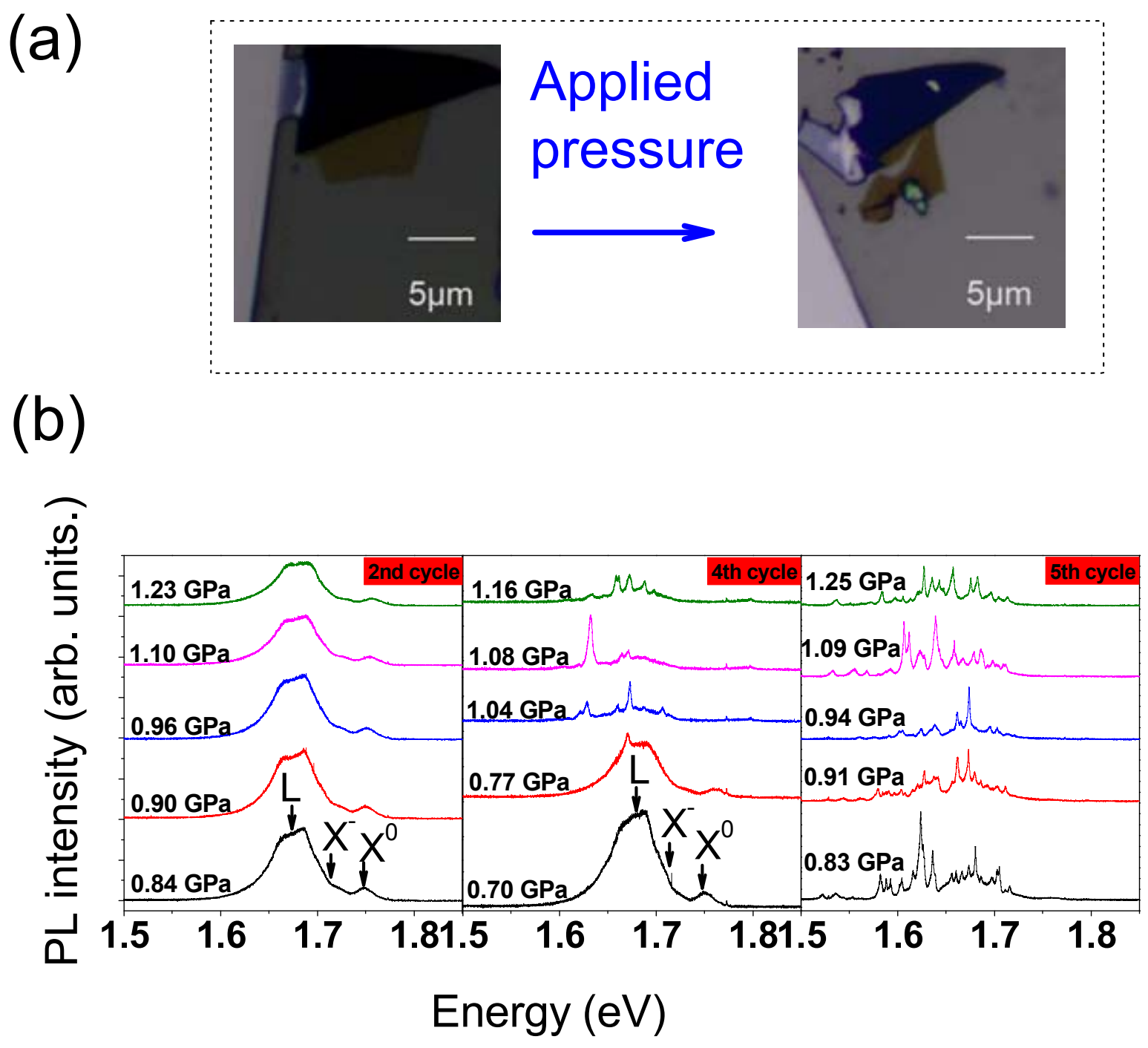

FIG. 4. (a) Optical microscope images obtained before pressure applying (left) and after releasing the pressure (right). (b) PL spectral evolution as a response to the pressure for the 2nd, 4th and 5th period of pressure-tuning cycles obtained by *in situ* tuning the pressure through piezoelectric actuator.

## REFERENCES

[1]A. Srivastava, M. Sidler, A. V. Allain, D. S. Lembke, A. Kis and A. Imamoglu, Nature nanotechnology **6**, 491 (2015).
[2]M. Koperski, K. Nogajewski, A. Arora, V. Cherkez, P. Mallet, J. Y. Veuillen, J. Marcus, P. Kossacki and M. Potemski, Nature nanotechnology **6**, 503 (2015).
[3]Y. M. He, G. Clark, J. R. Schaibley, Y. He, M. C. Chen, Y. J. Wei, X. Ding, Q. Zhang, W. Yao, X. Xu, C. Y. Lu and J. W. Pan, Nature nanotechnology **6**, 497 (2015).
[4]S. Kumar, M. Brotóns-Gisbert, R. Al-Khuzheyri, A. Branny, G. Ballesteros-Garcia, J. F. Sánchez-Royo and B. D. Gerardot, Optica **8**, 882 (2016).

[5]P. Tonndorf, R. Schmidt, R. Schneider, J. Kern, M. Buscema, G. A. Steele, A. Castellanos-Gomez, H. S. J. van der Zant, S. Michaelis de Vasconcellos and R. Bratschitsch, Optica **4**, 347 (2015).
[6]C. Chakraborty, L. Kinnischtzke, K. M. Goodfellow, R. Beams and A. N. Vamivakas, Nature nanotechnology **6**, 507 (2015).
[7]K. F. Mak and J. Shan, Nature Photonics **4**, 216 (2016).
[8]T. T. Tran, K. Bray, M. J. Ford, M. Toth and I. Aharonovich, Nature nanotechnology **1**, 37 (2016).
[9]A. Branny, G. Wang, S. Kumar, C. Robert, B. Lassagne, X. Marie, B. D. Gerardot and B. Urbaszek, Appl Phys Lett **14** (2016).
[10]S. Kumar, A. Kaczmarczyk and B. D. Gerardot, Nano Lett **11**, 7567 (2015).
[11]J. Kern, I. Niehues, P. Tonndorf, R. Schmidt, D. Wigger, R. Schneider, T. Stiehm, S. Michaelis de Vasconcellos, D. E. Reiter, T. Kuhn and R. Bratschitsch, Adv Mater **33**, 7101 (2016).
[12]X. F. Wu, X. M. Dou, K. Ding, P. Y. Zhou, H. Q. Ni, Z. C. Niu, D. S. Jiang and B. Q. Sun, Applied Physics Letters **25** (2013).
[13]A. M. Jones, H. Yu, N. J. Ghimire, S. Wu, G. Aivazian, J. S. Ross, B. Zhao, J. Yan, D. G. Mandrus, D. Xiao, W. Yao and X. Xu, Nature nanotechnology **9**, 634 (2013).
[14]C. R. Zhu, K. Zhang, M. Glazov, B. Urbaszek, T. Amand, Z. W. Ji, B. L. Liu and X. Marie, Physical Review B **16**, 161302 (2014).
[15]A. Srivastava, M. Sidler, A. V. Allain, D. S. Lembke, A. Kis and A. Imamoğlu, Nature Physics **2**, 141 (2015).
[16]G. Wang, L. Bouet, D. Lagarde, M. Vidal, A. Balocchi, T. Amand, X. Marie and B. Urbaszek, Physical Review B **7** (2014).
[17]Y. You, X.-X. Zhang, T. C. Berkelbach, M. S. Hybertsen, D. R. Reichman and T. F. Heinz, Nature Physics **6**, 477 (2015).
[18]C.-H. Chang, X. Fan, S.-H. Lin and J.-L. Kuo, Physical Review B **19**, 195420 (2013).
[19]M. Pena-Alvarez, E. del Corro, A. Morales-Garcia, L. Kavan, M. Kalbac and O. Frank, Nano letters **5**, 3139 (2015).
[20]S. B. Desai, G. Seol, J. S. Kang, H. Fang, C. Battaglia, R. Kapadia, J. W. Ager, J. Guo and A. Javey, Nano letters **8**, 4592 (2014).
[21]C. R. Zhu, G. Wang, B. L. Liu, X. Marie, X. F. Qiao, X. Zhang, X. X. Wu, H. Fan, P. H. Tan, T. Amand and B. Urbaszek, Physical Review B **12** (2013).
[22]Y. Ye, X. Dou, K. Ding, D. Jiang, F. Yang and B. Sun, Nanoscale **20**, 10843 (2016).
[23]B. H. Cheong and K. J. Chang, Phys Rev Lett **26**, 4354 (1993).
[24]W. P. Roach, M. Chandrasekhar, H. R. Chandrasekhar, F. A. Chambers and J. M. Meese, Semicond Sci Tech **4**, 290 (1989).
[25]H. Y. Nan, Z. L. Wang, W. H. Wang, Z. Liang, Y. Lu, Q. Chen, D. W. He, P. H. Tan, F. Miao, X. R. Wang, J. L. Wang and Z. H. Ni, Acs Nano **6**, 5738 (2014).